\documentclass[aps,pre,twocolumn,superscriptaddress]{revtex4-1}
\usepackage{amsmath,amsfonts,amssymb,wasysym,physics}
\usepackage{graphicx,color} % for pdflatex
\definecolor{myblue}{rgb}{0.368417, 0.506779, 0.709798}
\definecolor{myyellow}{rgb}{0.880722, 0.611041, 0.142051}
\definecolor{mygreen}{rgb}{0.560181, 0.691569, 0.194885}
\definecolor{myred}{rgb}{0.922526, 0.385626, 0.209179}
\begin{document}
\title{
Changes of graph structure of transition probability matrices\\ indicate the slowest kinetic relaxations
% Graph structure changes of transition probability matrices\\
% indicate
% the slowest kinetic relaxations
}
\author{Teruaki Okushima}
\email{okushima@isc.chubu.ac.jp}
\affiliation{
College of Engineering, Chubu University, Matsumoto-cho, Kasugai, Aichi 487-8501, Japan}
\author{Tomoaki Niiyama}
\email{niyama@se.kanazawa-u.ac.jp}
\affiliation{
College of Science and Engineering,
Kanazawa University,
Kakuma-cho, Kanazawa, Ishikawa 920-1192, Japan
}
\author{Kensuke S. Ikeda}
\email{ahoo@ike-dyn.ritsumei.ac.jp}
\affiliation{
College of Science and Engineering, Ritsumeikan University, 
Noji-Higashi 1-1-1, Kusatsu 525-8577, Japan}
\author{Yasushi Shimizu}
\email{shimizu@se.ritsumei.ac.jp}
\affiliation{Department of Physics, Ritsumeikan University, Noji-Higashi 1-1-1, Kusatsu 525-8577, Japan}
\date{\today}
\begin{abstract} 
Graphs of the most probable transitions for 
a transition probability matrix, $e^{\tau K}$, 
 i.e., the time evolution matrix of the transition rate matrix $K$
 % that is the time evolution matrix of the transition rate matrix $K$
over a finite time interval $\tau$,
are considered.
 We study  how the graph structures of the  most probable transitions
 change as  functions of $\tau$,
thereby elucidating that 
a kinetic threshold $\tau_g$ for the  graph structures exists.
Namely,
for $\tau<\tau_g$,
the number of connected graph components are constant.
In contrast,
for $\tau\geqslant \tau_g$,
recombinations of most probable transitions over the connected
graph components occur multiple times,
which introduce  drastic changes into the graph structures.
Using an illustrative  multi-funnel model,
we show that
 the recombination patterns indicate
 the existence of
the eigenvalues and eigenvectors of slowest relaxation modes
 quite precisely.
{We also devise 
 an evaluation formula %for correcting eigenvalues
that enables us 
 to  correct  the values of eigenvalues  with high accuracy
 from the data of merging processes.
 We show that  the graph-based method %developed in this paper
 is valid for a wide range of kinetic systems
 with  degenerate,
 as well as non-degenerate, relaxation rates.}
\end{abstract}
\maketitle

\section{introduction}

Complex relaxation dynamics,
such as
glass dynamics \cite{goldstein,stillingerWeber,stillingerWeber2,stillinger,heuer,APRV,debenedettiStillinger,sastory,DRB,DH,DH2,AFMK,glass,DSSSKG,yang,doliwaheuer2},
folding of biomolecules
\cite{BeckerKerplus,wolynes,pande,wang,folding,kern,nucleosome,BPE, hummer,hummer2,hummer3}, and
microcluster dynamics 
\cite{bbskj,hhdksi,jkk,hshias,hbssh,niiyama,kimura,cubic,maeno,oku2007},
are 
described 
frequently by kinetic differential equations
\cite{wales0,wales,stillingerBook}:
\begin{equation}
\frac{d}{dt} \vb*{p}=K \vb*{p},
\label{eq:kineticEq}
\end{equation}
where
$\vb*{p}=(p_1,\dots, p_i,\dots,p_n)$
is
the probability distribution vector,
and
$p_i$ is 
the probability of being in  state $i$,
with $n$ denoting the number of states.
$K$ is 
the transition rate matrix,
whose off-diagonal   $(i,j)$ element
describes 
the transition rate from the individual state $j$ to the other state $i$, 
and 
whose $j$th  diagonal element is chosen to %so as to  
satisfy 
$\sum_{i=1}^n (K)_{i,j}=0$ for  each $j$.
Then,
the total probability, $\sum_{i=1}^n p_i$, is conserved,
and,
under a general condition called ergodicity,
the eigenvalues, $\lambda_k$,  of $K$
satisfy the following relation \cite{haken}:
\[
\lambda_1=0 > \lambda_2
\geqslant \lambda_3 \geqslant \dots \geqslant \lambda_n.
\]
Moreover,
there exist corresponding eigenvectors $\vb*{v}_k$
%($k=1,2,\dots,n$)
such that
$\vb*{v}_1$ is the equilibrium of the system
satisfying
$\sum_{i=1}^n(\vb*{v}_1)_i=1$,
and
$\vb*{v}_k$ ($k=2,3,\dots$) are the relaxation modes
satisfying
$\sum_{i=1}^n(\vb*{v}_k)_i=0$.

On the other hand, %However,
it is a transition probability matrix $T(\tau)$,
describing the time evolution over a finite time interval $\tau$,
that
is observed experimentally \cite{BPE}.
The mapping of the  time evolution is given by
\begin{equation}
 \vb*{p}(t+\tau)=T(\tau) \vb*{p}(t)
\label{eq:kineticMap}
\end{equation}
for $t=0,\ \tau,\ 2\tau,\ \dots$,
where $T(\tau)=\exp(\tau K)$ holds 
for the  system with the transition rate matrix of $K$.
The eigenvalues of $T(\tau)$
are given by  $\bar\lambda_k(\tau)=e^{\tau \lambda_k}$
and hence
satisfy the following relation 
\cite{BPE}:
\begin{equation}
\bar\lambda_1(\tau)=1 > \bar\lambda_2(\tau)
\geqslant \bar\lambda_3(\tau)\geqslant \dots \geqslant\bar\lambda_n(\tau)>0. 
\end{equation}
Note here that
$\tau$ is regarded as 
a coarse-graining parameter of time,
because
any relaxation modes, relaxing with rates  faster than $1/\tau$,
satisfy
$\bar\lambda_k(\tau)\sim 0$
and hence are effectively neglected from $T(\tau)$.

{
Equations
\eqref{eq:kineticEq}
and 
\eqref{eq:kineticMap}
are  called 
the continuous-time and  discrete-time Markov state models,
respectively.
These models
have been studied extensively 
 in a wide range of  fields.
% Especially,
In particular,
 Markov properties,
in which
all transitions  from  an arbitrary state
do not depend on any  previous states,
have been studied in detail
because the properties are conditions for satisfying Markov state models.
For example,
how one can introduce coarse-grained states that
 ensure the Markov property was  studied in \cite{yang}.
 From
 the $\tau$-dependence of  the eigenvalues of $T(\tau)$,
 the conditions of $\tau$ to
 ensure the Markov properties were elucidated in \cite{hummer}.
 In addition to  these works, for the renormalization problem,
 i.e., studies on
how to derive lower-dimensional effective Markov state models,
a technique with use of  the Perron cluster algorithm
was invented in \cite{hummer2}.
 (See, for a review, \cite{BPE}.)
}
Also, we have developed
a renormalization method for the Markov state models, %Eq.~\eqref{eq:kineticEq},
where
renormalized transition rates 
between 
 coarse-grained states, called metabasins \cite{kfs}, are defined.
We have shown there  that
the slowest relaxations are obtained accurately
with this method \cite{rg}.

In this paper, 
by using a multi-funnel model that  has been used
in our previous studies in Refs.~\cite{maeno,rg},
we demonstrate that the metabasin analysis,
based upon the most probable path graphs,
is successfully  applied to
the kinetic differential equations of Eq.~\eqref{eq:kineticEq}
in Sec.~\ref{sec:K}
and 
the coarse-grained time maps %(or the time-domain coarse-grained map)
of Eq.~\eqref{eq:kineticMap}
in Sec.~\ref{sec:graphT}.
We then show that
these graphs
 can describe illustratively the characteristics  
of multi-timescale relaxation dynamics.
In particular, 
a characteristic threshold time $\tau_g$,
at which the intra-funnel relaxation dynamics switches
to the inter-funnel relaxation dynamics, 
is definitely extracted.  
More specifically,
we will see that
for $\tau<\tau_g$,
the metabasins of $T(\tau)$ 
are composed of almost the same states,
in spite of the very frequent intra-metabasin recombinations
of most probable transitions.
In contrast,
for $\tau\geqslant\tau_g$
the metabasins begin to merge with each other,
due to the inter-metabasin recombinations of most probable transitions.
We then elucidate
in Sec.~\ref{subsubIntutive},
how and why
these 
graph structure changes
correspond
to
the eigenvalues and the eigenvectors of the slowest relaxation modes.
{Furthermore,
in Sec.~\ref{hosei},
we  devise
an evaluation formula
that enables us to correct the values of the eigenvalues
with high accuracy from the data of merging processes.
As shown in Sec.~\ref{subsubDegenerate},
these graph-based methods are valid for the degenerate,
as well as non-degenerate, relaxation rate systems.}

{
% To the best of the authors' knowledge,
% this is the first study that elucidates
In this study,
we elucidate
how one can extract
information about relaxation rates and eigenvectors of $K$
from the graph structures of $T(\tau)$.
}

\section{model}

In this section,
we introduce  the four-funnel model used in Refs.~\cite{maeno,rg},
which models basin hopping on 
high-dimensional potential energy landscapes.
We  assume 
that
the intra-basin relaxation modes
relax so fast 
that
any probability densities $\rho(\vb*{r})$,
where $\vb*{r}$ is a coordinate vector of all atoms,
are 
expressed as
the  linear combinations of 
the intra-basin local equilibria, $\rho_i(\vb*{r})$,
in basins of $i$.
Namely,
$\rho(\vb*{r})=\sum_{i=1}^n p_i \rho_i(\vb*{r})$ holds,
where $n$ is the number of basins
and
$p_i$ is 
the probability
of being in
basin $i$.
Hence,
the probability density $\rho(\vb*{r})$ is
fully specified
by
the probability vector
$\vb*{p}=(p_1,\dots, p_i,\dots,p_n)$.
In addition,
the saddle point, which connects basins $i$ and $j$, is denoted by $ij$,
so that  $ij=ji$ holds.
Moreover, we assume that
the hopping rates between
the adjacent basins %$i$ and  $j$
are given by the 
Arrhenius transition rates \cite{wales,stillingerBook}.
Namely,
the transition rate from state $j$ to state $i(\neq j)$
is
given by
\begin{equation}
k_{i,j}=\nu_{i,j} e^{-\beta(E_{ij}-E_j) }, 
\label{eq:Arrhenius}
\end{equation}
where
$E_{ij}$ is the energy of saddle point $ij$,
$E_j$ is the minimum energy in basin $j$,
and
$\nu_{i,j}$
is the frequency factor of 
transition $j\to i$. 
In the following,
we set $\nu_{i,j}=1$ for all $i$ and $j$,
for the sake of simplicity.

\begin{figure}[t]
\includegraphics[width=6cm]{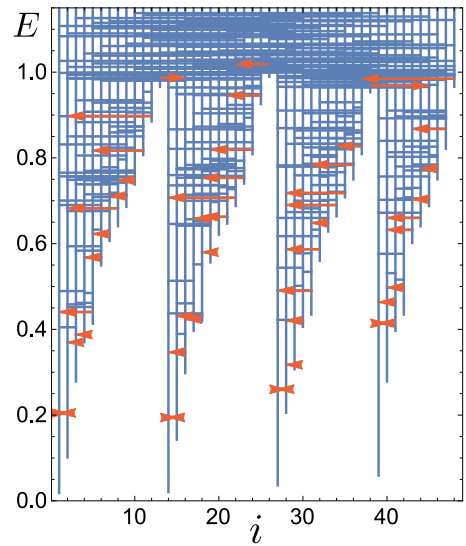}
\caption{%
Saddle connectivity graph for 
a four-funnel model \cite{maeno}.
Vertical upward lines at $i=1,2,\dots,48$ starting at $E_i$
represent states of $i$.
Horizontal lines at $E=E_{ij}$ connecting $i$ and $j$
represent saddles $ij$.
Most probable transitions $j\to i$
are also shown by
(red) arrows from $j$ to $i$.
(See the text)
\label{fig:scg}
} 
 \end{figure}

Figure \ref{fig:scg} 
depicts a four-funnel model
in  what we call a saddle connectivity graph,
where
basin energies $E_i$
and saddle energies $E_{ij}$
are represented
for all basins of  $i$ 
and for all saddles of $ij$, respectively \cite{maeno}.
For the sake of reproducibility,
 the supplemental materials
of 
LM4funnel.dat %\cite{sm1}
and
SP4funnel.dat \cite{sm1}
are attached to this paper.
$E_i$ is written  in the $i$th line of LM4funnel.dat.
The triplet data of $i$, $j$, and  $E_{ij}$ are written in each line of
SP4funnel.dat.
Figure \ref{fig:scg} shows that
metabasins of 
MB$_1$,
MB$_2$,
MB$_3$, and
MB$_4$
are composed of
$\{1,\dots,13\}$,
$\{14,\dots,26\}$,
$\{27,\dots,38\}$, and
$\{39,\dots,48\}$,
respectively,
where
the saddles densely
connect 
every state to
the other states
belonging to
the same metabasins
as well as
the different metabasins.
%
% that are
% not only 
% within the same metabasins
% but also 
% belonging to
% the different metabasins.
%
The shapes of metabasins
are said to be funnel-like,
because, in each metabasin,
there exist pathways
along which
the basin energies $E_i$
and
the saddle point energies $E_{ij}$,
respectively,
decrease
monotonically
upon
approaching
the minimum energy states in the metabasins.

 \begin{figure}[t]
 \includegraphics[width=6cm]{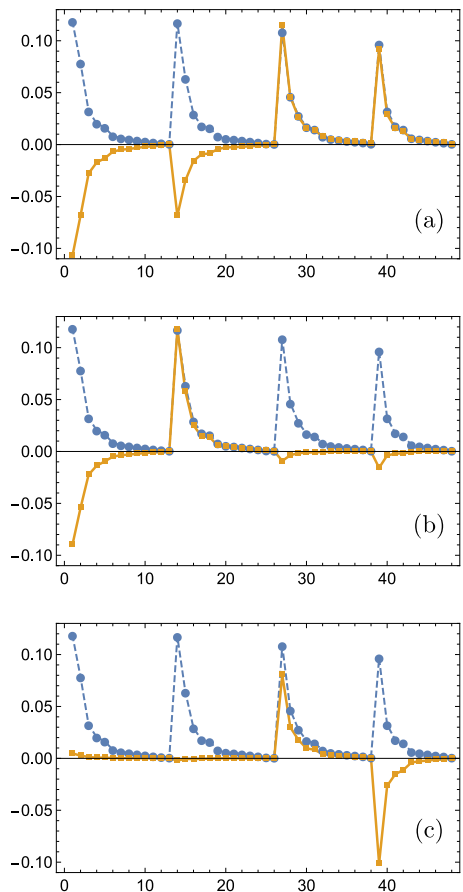}  
\caption{%
Eigenvectors  $\vb*{v}_2$, $\vb*{v}_3$, and $\vb*{v}_4$ 
of transition rate matrix $K$ for the four-funnel model of Fig.~\ref{fig:scg}
at $\beta=5$ are plotted in (a), (b), and (c), 
respectively. 
In each plot,
the equilibrium, $\vb*{v}_1$,
is also plotted using circles connected with a dashed line for comparison.
Here, 
$\vb*{v}_1$ is normalized as
$\sum_{i=1}^n(\vb*{v}_1)_i=1$,
and
$\vb*{v}_k$ ($k\geqslant 2$) are scaled 
such that
the components satisfying $(\vb*{v}_k)_i>0$ approximately agree with $(\vb*{v}_1)_i$.
\label{fig:evec}
} 
 \end{figure}

With the use of Eq.~\eqref{eq:Arrhenius},
we calculated $k_{i,j}$ at $\beta=5$
and
diagonalized  matrix $K$,
 whose off-diagonal elements are given by $(K)_{i,j}=k_{i,j}$.
The eigenvalues of
$\lambda_1$, $\lambda_2$, $\lambda_3$, $\lambda_4,\dots$
are given by
\begin{align}
0, -0.0886, -0.154, -0.235, -1.285, \dots,
\label{eq:eig}
\end{align}
where
$\lambda_1=0$ 
corresponds to the  equilibrium, % at $\beta=5$,
$\lambda_2$, $\lambda_3$, $\lambda_4 \sim\!-0.1$
are 
the three slowest relaxation rates,
and
$\lambda_k < -1  
$ ($k\geqslant 5$)
are  the relaxation rates that are
more than  one order of magnitude
faster than the slowest relaxation of $\lambda_2$.
In Fig.~\ref{fig:evec},
we show 
these slowest  eigenvectors $\vb*{v}_k$ for $k=1,\dots,4$.
We see that
the equilibrium distribution $\vb*{v}_1$
(circles with dashed lines in Fig.~\ref{fig:evec})
is 
the superposition of 
the four intra-metabasin local equilibrium distributions,
which
have 
the local maximal probabilities 
at the funnel bottoms of $i=1,14,27,39$.
Figure \ref{fig:evec}(a)
shows that
the slowest relaxation mode $\vb*{v}_2$
generates
the 
probability flow from
the local equilibrium in $\{27,\dots,48\}$
to
the local equilibrium in $\{1,\dots,26\}$.
Similarly,
$\vb*{v}_3$
generates
%corresponds to
the probability flow
from the local equilibrium in $\{14,\dots,26\}$
to that  in $\{1,\dots,13\}$ [Fig.~\ref{fig:evec}(b)]
and
$\vb*{v}_4$
generates the probability flow
from that in $\{27,\dots,38\}$
to that in $\{39,\dots,48\}$
[Fig.~\ref{fig:evec}(c)].
(For details, see the discussion in Sec.~\ref{sec:gt_tau_g}.)

{
Note here that
there are various ways of introducing metabasins.
For example,
Perron cluster algorithms 
utilize  the slowest relaxation eigenvectors of Markov state models \cite{BPE},
and
other lumping methods 
combine states that are separated by small energy barriers \cite{doliwaheuer2}.
In the following,
the metabasins
are introduced 
with the use of the most probable transitions,
in the same manner as in \cite{oku2007,yang}.
}

\section{Most probable path graph of $K$\label{sec:K}}
\begin{figure}[t]
\includegraphics[width=6cm]{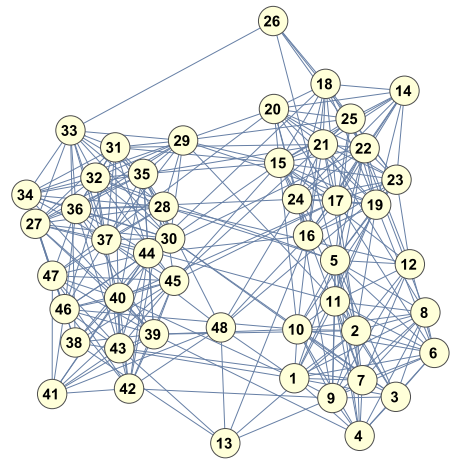}
\caption{%
Transition graph for the four-funnel model of Fig.~\ref{fig:scg},
where all possible transitions between the saddles of $ij$
are represented by
the edges connecting vertices $i$ and $j$.
\label{fig:GofK}} 
\end{figure}

 \begin{figure*}[t]
 \includegraphics[width=16cm]{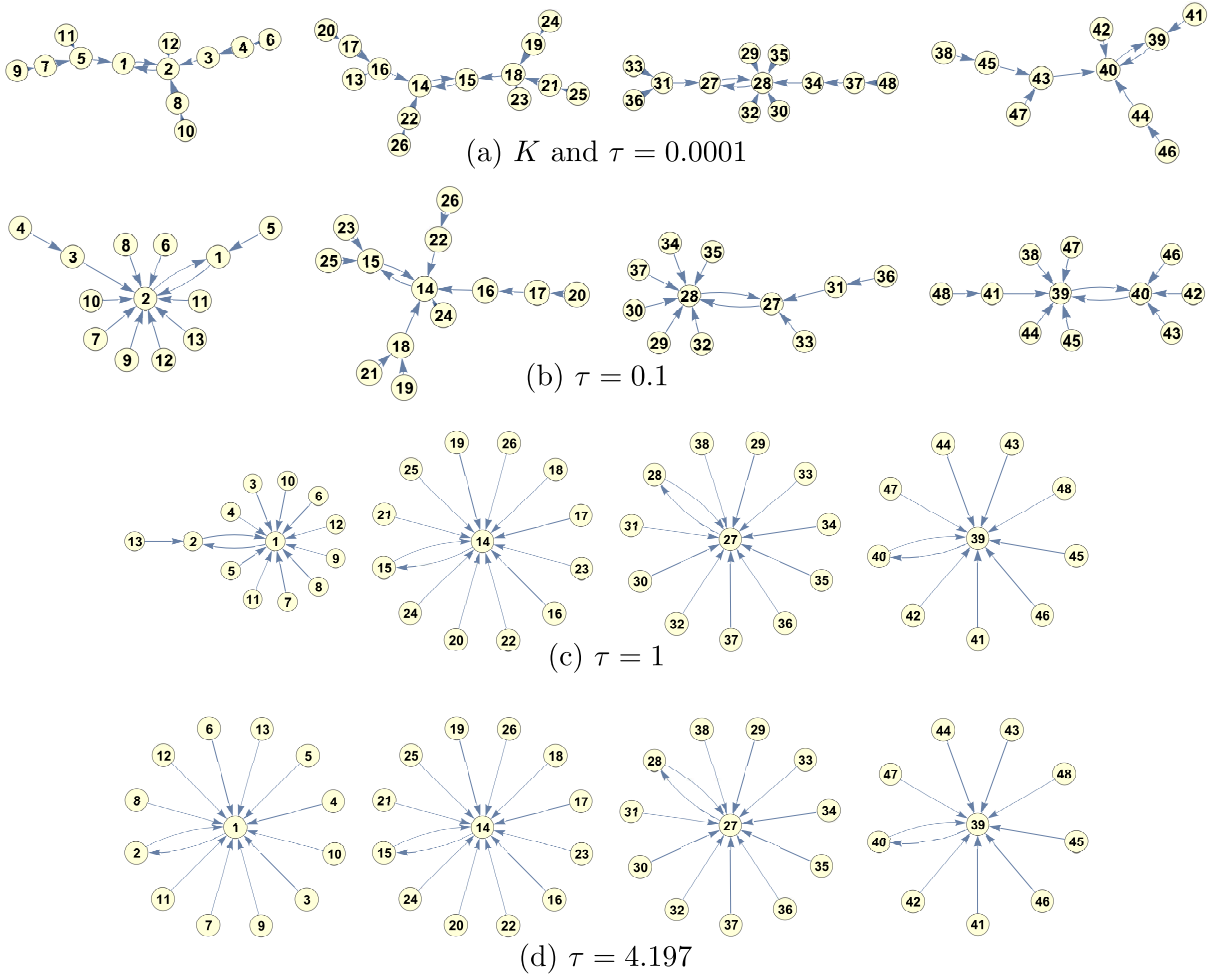}
\caption{%
 Most probable path graphs
for (a) $K$ and $T(\tau)$ with $\tau=0.0001$
(both are the same graph)
and 
for $T(\tau)$ with
(b) $\tau=0.1$,
(c) $\tau=1$,
(d) $\tau=4.197$.
Each graph of (a)--(d) has
four connected graph components,
which we call the metabasin.
(See the text)
\label{fig:tau0.0001}
} 
 \end{figure*}

In this section,
we discuss
why and how 
we introduce the most probable path graph.

Saddles $ij$($=ji$)
enable the  transitions of $i\to j$ and $j\to i$.
Hence,
we can draw 
a graph by connecting  the indices of states  $i$ and $j$ by edges,
for all saddles of $ij$.
Figure~\ref{fig:GofK} shows the transition graph of the four-funnel
model depicted in Fig.~\ref{fig:scg}.
The graph contains
all information about 
possible transitions or basin adjacencies,
except the information about energy levels of $E_i$ and  $E_{ij}$ 
due to the contraction of energy-height information.
However,
the funnel structures,
which are seen in the saddle connectivity graph of Fig.~\ref{fig:scg},
are not apparent %clear %buried away
in Fig.~\ref{fig:GofK},
due to
the cumbersome graph structure. 
The reasons for the failure to capture the funnel features
is because
the 
important transitions
and the unimportant
transitions
are equally drawn in Fig.~\ref{fig:GofK}.

To tame the  graph structural complexity,
we here introduce an alternative graph that 
consists only of 
the most important transitions.
Suppose that
the probability vector $(p_1,p_2,\dots,p_n)$
at a moment is  given by
$p_i=\delta_{i,j}$. % ($i=1,2,\dots, n$). 
Then,
any transitions $j\to i$ 
can occur at the moment, if $k_{i,j}\neq 0$ ($i\neq j$). 
% any transitions  to states $i$ of $k_{i,j}\neq 0$ with $i\neq j$
% can occur
% at the moment.
Hence,
the most probable transition from $j$
is given by
$j\to i$
such that 
$k_{i,j} =\max\{ k_{i',j}\mid
1\leqslant i' \leqslant n,\  i' \neq j
\}$.
%
% satisfying
% $k_{i,j}\geqslant k_{i',j}$ for 
% all $i'$
% with  $1\leqslant i' \leqslant n$  ($ i' \neq j$).
%
In Fig.~\ref{fig:scg},
all of the most probable transitions $j\to i$ 
are shown by red arrows for the four-funnel model,
from which we see
that
the most probable transitions 
 are folded in the four funnels.
This means that 
the funnel structures can be extracted
by the simpler subgraph of  the most probable transitions.

In Fig.~\ref{fig:tau0.0001}(a),
we show
the most probable path graph of $K$,
which is the directed graph of most probable transitions for $K$,
where
all the most probable transitions $j$ to $i$
are represented by the arrows from $j$ to $i$.
As we expected,
the graph is composed of the four connected graph components,
which 
correspond to the four metabasins %, or funnels, 
of MB$_1$, MB$_2$, MB$_3$, and MB$_4$ depicted by red arrows
 in Fig.~\ref{fig:scg}. 
Moreover,
we see that
 each graph component
has
an attracting cycle $i\to j \to i$ 
(in the following abbreviated as $i \Leftrightarrow j$ for simplicity) 
containing  the lowest energy state in the corresponding metabasin.
Hence, we introduce the following compact notation:
\begin{align}
\begin{aligned}
& \text{MB}_1(1\Leftrightarrow 2)
=\{1, 2, \dots,  12\},\\
& \text{MB}_2(14\Leftrightarrow 15)
=\{13, 14, \dots,  26\},\\
& \text{MB}_3(27\Leftrightarrow 28)
=\{27, 28, \dots, 37, 48\},\\
& \text{MB}_4(39\Leftrightarrow 40)
=\{38, 39, \dots, 47\},
\end{aligned}
\label{eq:mb0}
\end{align}
where
MB$_k(i\Leftrightarrow j)=\{i, j, j', j'', \dots\}$
means that
MB$_k$ with cycle $i\Leftrightarrow j$ is composed  of
$\{i, j, j', j'', \dots\}$.

In this section,
we have confirmed
that
the most probable path graph [Fig.~\ref{fig:tau0.0001}(a)],
as well as the saddle connectivity graph (Fig.~\ref{fig:scg}),
can extract the metabasin structures of  transition rate matrices of $K$.

\section{Most probable path graph of $T(\tau)$\label{sec:graphT}}
Unfortunately,
the saddle connectivity graph,
{as well as other graphing methods, such as
the disconnectivity graph \cite{wales}},
is not applicable to
the transition probability matrix $T(\tau)$,
because 
both $E_j$ and $E_{ij}$,
which are  indispensable for
drawing these graphs,
 are not defined in $T(\tau)$.
In contrast,
the most probable path graph 
of $T(\tau)$
is naturally defined, as shown below.

The transition probability  from $j$ to $i$ in the duration of time $\tau$
is given by $(T(\tau))_{i,j}$.
Hence,
the most probable transition from $j$ in $\tau$
is given by
$j\to i$
such that
$(T(\tau))_{i,j}=\max\{ (T(\tau))_{i',j}\mid
1\leqslant i' \leqslant n,\quad  i'\neq j
\}$.
%
% satisfying
% $(T(\tau))_{i,j}\geqslant (T(\tau))_{i',j}$
% for $1\leqslant i' \leqslant n$ ($ i'\neq j$).
%
The most probable path graph of $T(\tau)$
is drawn by arrows from  $j$ to $i$ for
all of the most probable transitions $j\to i$ without difficulty,
 in the same way as the graph for $K$ was drawn.

In the following,
we study
the structural changes of 
the most probable path graph of  $T(\tau)$ with varying $\tau$,
thereby elucidating that
there exists a kinetic threshold, $\tau_g$,  of time interval such that
\begin{equation}
\tau_g \simeq 4.198.
% \tau_g = 4.1979\cdots.
\end{equation}
Specifically,
the members of metabasins are approximately conserved  for $\tau<\tau_g$
(Sec.~\ref{sec:lt_tau_g}),
while
there are several mergings of metabasins for $\tau\geqslant\tau_g$
(Sec.~\ref{sec:gt_tau_g}).

{
Also in Ref.~\cite{yang},
the most probable paths of  $T(\tau)$
were studied to investigate the  Markov property in metabasin space
for a glass former.
Moreover, in Ref.~\cite{hummer},
the $\tau$-dependencies of $T(\tau)$
were elucidated in order to examine
the Markov property in the eigenvector space.
% Remark that
% the $\tau$-dependencies of
% the most probable path graphs are examined for the first time in this study.
}

\subsection{ $\tau < \tau_g$ case\label{sec:lt_tau_g}}
As shown in Figs.~\ref{fig:tau0.0001}(a)--\ref{fig:tau0.0001}(d),
 the most probable path graphs of $T(\tau)$ for $\tau < \tau_g$
have four connected graph components.
At $\tau=0.0001$,
the most probable path graphs %of most probable transitions
for $T(\tau)$ and $K$
are identical.
Hence,
we call the four connected components 
the metabasins of MB$_1$, MB$_2$, MB$_3$, and MB$_4$ for $T(\tau)$,
as we did for $K$ in Sec.~\ref{sec:K}.
Of course,
all the cycles and members of MB$_k$
are the same as Eq.~\eqref{eq:mb0}.
Note here that
the coincidences of metabasins of $K$ and $T(0.0001)$
means that
the metabasin coarse-graining 
developed in Ref.~\cite{rg}
is a  sound basis 
for the stable description, or the renormalization, of the kinetic evolutions
of Eqs.~\eqref{eq:kineticEq} and \eqref{eq:kineticMap}.

\begin{figure*}[t]
 \includegraphics[width=11.5cm]{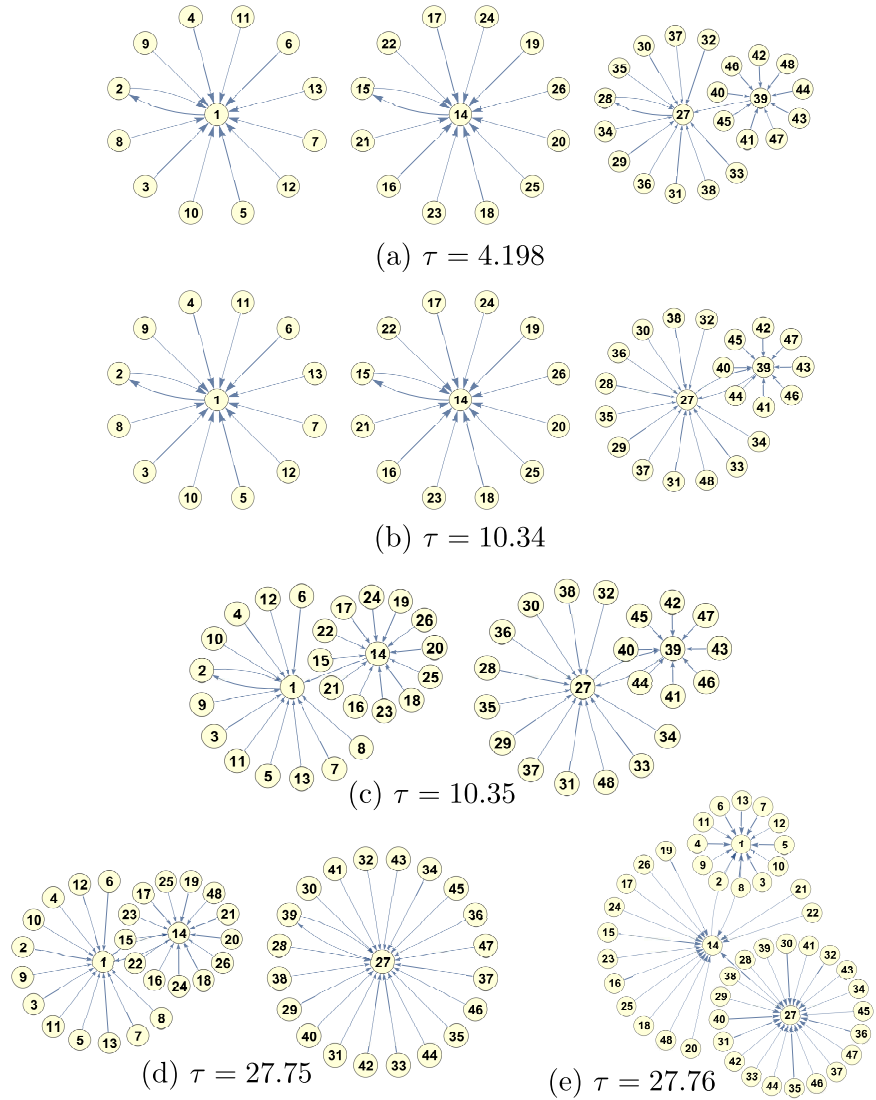}
\caption{%
 Most probable path graphs of $T(\tau)$
 for
 (a) $\tau=4.198$,
 (b) $\tau=10.34$,
 (c) $\tau=10.35$,
 (d) $\tau=27.75$,
 and
 (e) $\tau=27.76$,
which 
show the important
recombinations of most probable transitions
that lead the merging processes of metabasins.
\label{fig:tau4.2}
} 
\end{figure*}

The most probable path graph of $T(\tau)$ with $\tau=0. 1$
is shown in Fig.~\ref{fig:tau0.0001}(b),
where, 
while all of the cycles are the same as 
the cycles
of $\tau=0.0001$ in Fig.~\ref{fig:tau0.0001}(a),
the members of metabasins slightly change:
state $13$ moves from  MB$_2$ to MB$_1$
and 
state $48$ moves from MB$_3$ to MB$_4$,
which results in
\begin{align}
\begin{aligned}
& \text{MB}_1(1\Leftrightarrow 2)
=\{1, 2, \dots,13\},\\
& \text{MB}_2(14\Leftrightarrow 15)
=\{14,\dots,26\},\\
& \text{MB}_3(27\Leftrightarrow 28)
=\{27, 28, \dots, 37\},\\
& \text{MB}_4(39\Leftrightarrow 40)
=\{38, 39, \dots, 47, 48\}.
\end{aligned}
\label{eq:mb1}
\end{align}
Note here that
the moving states of $13$ and $48$
are the peripheral states
that
are far from the cycles.

More specifically,
the graph  of Fig.~\ref{fig:tau0.0001}(b)
consists of  
the more direct transitions to the attractive cycles 
compared to Fig.~\ref{fig:tau0.0001}(a).
Hence,
the longer the time evolution is,
the more directly
the states arrive at the attractive cycles,
which means that
the probability vectors from any states
tend to  evolve into the intra-MB local equilibria
to which they belong.

For $1\leqslant \tau < \tau_g$, 
the compositions and the cycles of metabasins do not change as
\begin{align}
\begin{aligned}
& \text{MB}_1(1\Leftrightarrow 2)
=\{1,  \dots,  13\},\\
& \text{MB}_2(14\Leftrightarrow 15)
=\{14, \dots, 26\},\\
& \text{MB}_3(27\Leftrightarrow 28)
=\{27, \dots, 38\},\\
& \text{MB}_4(39\Leftrightarrow 40)
=\{39, \dots,  48\}.
\end{aligned}
\end{align}
Note here that
these  MBs agree with the MBs read from  Fig.~\ref{fig:scg}.
Moreover,
comparing %the graphs (c) and (d) in 
Figs.~\ref{fig:tau0.0001}(c) and \ref{fig:tau0.0001}(d),
we see that
the most probable transitions 
are reconnected within MB$_1$
in the time duration 
from 
 $\tau=1$
to
$\tau=4.197$.
Especially at $\tau=4.197$,
all transitions become the direct transitions
to 
the most probable states in the intra-metabasin local equilibria,
which means that
all states within a metabasin 
evolve 
to the intra-metabasin local equilibrium
in the course of time with  $\tau=4.19$,
and, as a result,
that
all the most probable transitions
become the direct transitions to 
the most probable states in the intra-metabasin local equilibria.

In this subsection,
we elucidated the following:
(a) The members of metabasins of $T(\tau)$ 
remain almost unchanged 
for $\tau <\tau_g$.
(b) The cycles of metabasins remain exactly the same.
(c) The moving peripheral states
 that are located far from the attracting cycles
 can  change  metabasins to which they belong.
(d)
At $\tau=4.19$,
all states evolve
to the intra-MB local equilibria,
and thus 
the most probable transitions at the time become
the direct transitions to the lowest energy states
in the metabasins. 

\subsection{$\tau \geqslant \tau_g$ case\label{sec:gt_tau_g}}
For $\tau \geqslant \tau_g$,
the metabasins of $T(\tau)$ 
merge with each other several times,
as shown in Figs.~\ref{fig:tau4.2}(a)--\ref{fig:tau4.2}(e).
The merging processes are essentially
described by the recombinations of 
the transitions that are constituents of  attractive cycles.
As shown in Fig.~\ref{fig:tau0.0001}(d),
the metabasins  at $\tau=4.197$ ($<\tau_g$) include
\begin{equation}
\text{MB}_3(28\Leftrightarrow 27) \text{ and }
\text{MB}_4(39\Leftrightarrow 40).
\label{eq:mb3mb4}
\end{equation}
They merge with each other 
and 
form a bigger metabasin MB$_{\{3,4\}}$ at $\tau=4.198$ ($\geqslant\tau_g$), as
\begin{equation}
\text{MB}_{\{3,4\}}(28 \Leftrightarrow 27\leftarrow 39\leftarrow 40).
\label{eq:mb34}
\end{equation}
Expressions ~\eqref{eq:mb3mb4} and \eqref{eq:mb34}
clearly show that
the most probable transition $39 \to 40$  at $\tau=4.197$
is changed to $39\to 27$ at $\tau=4.198$.
% $27 \gets 39$ at $\tau=4.198$.
It is 
this newly created most probable transition
that is expected to induce the transport of
the excess probability between  MB$_3$ and  MB$_4$.
Let us confirm this expectation
by using the eigenvalues and eigenvectors of $K$.
First,
$\tau=4.198$ corresponds
to the rate of  $-1/\tau\sim -0.24$.
At around $-0.24$,
we indeed find the eigenvalue of $\lambda_4=-0.235$ 
from the list \eqref{eq:eig} of the eigenvalues. 
The corresponding  eigenvector $\vb*{v}_4$
is plotted in Fig.~\ref{fig:evec}(c),
where
the excess ($\vb*{v}_4>0$)
and the shortage ($\vb*{v}_4<0$)
from the equilibrium distribution,
respectively, correspond to the intra-MB$_3$ and intra-MB$_4$ local equilibria.
Both of these deviations change to zero as $\tau\to \infty$,
since
$\vb*{v}_4$ evolves as $T(\tau)\vb*{v}_4=e^{\lambda_4 \tau}\vb*{v}_4
=e^{-0.235\tau} \vb*{v}_4 \to 0$ ($\tau \to \infty$).
Hence,
as we expected,
the excess probability in the shape of the intra-MB$_3$ local equilibrium
is transported into the intra-MB$_4$ local equilibrium,
by the relaxation mode of $\vb*{v}_4$
at around $\tau=\tau_4=\tau_g= 4.198$.

The next merging of metabasins
occurs at  $\tau=\tau_3\equiv 10.35$.
Figure~\ref{fig:tau4.2}(b) %of $\tau=10.34$
shows that
the graph components of  MB$_1$ and MB$_2$ remain unchanged 
from $\tau=4.198$ [Fig.~\ref{fig:tau4.2}(a)],
while
the intra-metabasin structure
of 
MB$_{\{3,4\}}$ 
changes
from 
MB$_{\{3,4\}}(27\Leftrightarrow 28)$ 
to
MB$_{\{3,4\}}(27\Leftrightarrow 39)$,
which indicates
that
the intra-MB$_{\{3,4\}}$ local equilibrium
has been achieved until $\tau=10.34$.
(See the discussion in Sec.~\ref{subsubIntutive}.)
When $\tau$ becomes $\tau_3$,
MB$_1$($2\Leftrightarrow1$) and 
MB$_2$($14\Leftrightarrow 15$) merge with each other,
and the resulting metabasin
is 
MB$_{\{1,2\}} (2\Leftrightarrow 1 \leftarrow 14 \leftarrow 15)$.
Again,
we consider the eigenvalue and the eigenvector
corresponding to this merging process.
The eigenvalue that
corresponds to the rate of $-1/\tau_3 \sim -0.1$
is identified as $\lambda_3=-0.154$.
We plot 
the corresponding eigenvector $\vb*{v}_3$
in 
Fig.~\ref{fig:evec}(b),
which 
clearly shows that
the excess probability in the shape of the local equilibrium of MB$_2$
is transported 
to 
the local equilibrium of MB$_1$,
consistently with the graph merging process
at $\tau=\tau_3$.

Finally,
MB$_{\{1,2\}}(1\Leftrightarrow14)$
and
MB$_{\{3,4\}}(27\Leftrightarrow 39)$
merge with each other at
% a time %point
% from
%  $\tau=27.75$
% to
$\tau=\tau_2\equiv 27.76$,
and
the resulting metabasin is given by
\begin{equation}
 \text{MB}_{\{\{1,2\},\{3,4\}\}}
  (1\Leftrightarrow 14\leftarrow  27\leftarrow 39), 
\end{equation}
where we should point out that
MB$_{\{3,4\}}(27\Leftrightarrow 28)$ and 
MB$_{\{1,2\}}(1\Leftrightarrow 2)$ at $\tau=4.198$
have changed
to
MB$_{\{3,4\}}(27\Leftrightarrow 39)$ and 
MB$_{\{1,2\}}(1\Leftrightarrow 14)$
until $\tau=10.35$,
respectively.
These changes of cycles
mean 
the achievements of local equilibria
both 
in MB$_{\{1,2\}}$ and 
in MB$_{\{3,4\}}$.
From this merging process of MB$_{\{1,2\}}$ and MB$_{\{3,4\}}$,
we again expect 
the relaxation process
between MB$_{\{1,2\}}$  and MB$_{\{3,4\}}$
at around $-1/\tau_2 \sim -0.036$.
Let us confirm this expectation.
The corresponding eigenvalue
is $\lambda_2=-0.0886$,
and 
$\vb*{v}_2$ is plotted in 
Fig.~\ref{fig:evec}(a),
which clearly
shows
that
 the slowest relaxation mode of $\vb*{v}_2$  transports
the excess probability in the shape of the intra-MB$_{\{3,4\}}$ local equilibrium
to
the intra-MB$_{\{1,2\}}$ local equilibrium
at around the merging time $\tau_2 \sim 20.76$.

 \subsection{Summary\label{subsubSummary}}
Here,
we  summarize
the above findings of
how the graph structural changes
indicate
% correspond to 
the properties of  the eigenvalues and eigenvectors.

In $\tau<\tau_g$,
the most probable transitions 
are confined in MB$_k$ ($k=1,2,3,4$).
Hence,
only intra-metabasin equilibria can be achieved,
and 
the kinetic system remains  globally nonequilibrium. 
In contrast,
for $\tau \geqslant \tau_g$,
at around $\tau=\tau_2$, $\tau_3$, $\tau_4$ ($\tau_2>\tau_3>\tau_4=\tau_g$)
the most probable transitions between 
metabasins  are activated gradually. 
Especially for larger $\tau$,
the kinetic system
equilibrates globally
via 
the multiple graph structure changes of 
the most probable transitions.
Based on these findings,
we can say that
$\tau_g$ is a kind of glass-transition time
in a sense that
within  the activation time, 
the  inter-metabasin transitions are
effectively prohibited,
%while if $\tau$ is larger than $\tau_g$
while at $\tau=\tau_g$
the transitive phase-space volumes  become approximately doubled.
% while if it is longer than $\tau_g$
% the transitive phase space volumes  become to be approximately doubled.

\section{Discussion}
{
In this section,
we show in Sec.~\ref{subsubIntutive} that
the graph structural changes can be interpreted as
the manifestation of the time evolution of the local equilibria.
In Sec.~\ref{hosei},
we show
that
the discrepancies between merging rates $-1/\tau_i$
and the relaxation rates  $\lambda_i$
arise due to the lag times
from the beginnings of the relaxations
to the mergings of the basins.
Then, %In Sec.~\ref{hosei},
we derive
a formula for calculating $\lambda_i$
that corrects the errors arising from the lag times.
We also show that, with the formula,
one can evaluate the accurate  values of $\lambda_i$
from the actual merging process data.
Finally,
in Sec.~\ref{subsubDegenerate},
we consider 
the degenerate $\lambda_i$ case,
where
%and show that
the properties of the slowest relaxation modes
are 
derived 
similarly to the nondegenerate case,  by a graph-based analysis.
We confirm that also
in this degenerate case,
the correcting formula
produces  accurate estimates  of $\lambda_i$
from the actual merging process data.
}

\subsection{Intuitive explanation\label{subsubIntutive}}
\begin{figure}[t]
\includegraphics[width=5.5cm]{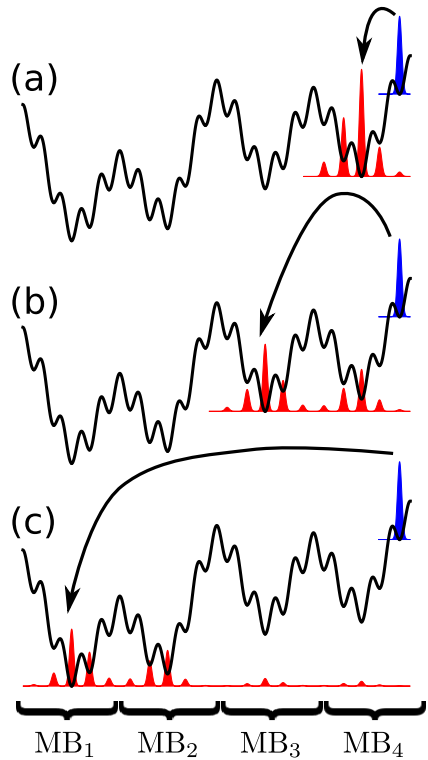}
\caption{%
Schematic illustration of 
the changes of most probable transitions as a function of $\tau$.
Suppose the initial 
probability distribution 
is the local equilibrium in  the rightmost basin,
which
is represented by the blue filled bell-shaped curve 
in each figure.
The most probable transitions from the rightmost state at various times of 
(a) $\tau=\tau_a$,
(b) $\tau_b$,
and 
(c) $\tau_c$ ($\tau_a<\tau_b<\tau_c$)
 are indicated by  curved  arrows.
%  in 
% (a), (b), and (c),
% respectively.
\label{fig:manga}} 
\end{figure}
Here,
we show 
that
it can be intuitively understood
why
the  structural changes of the most probable path graphs
indicate
the existence of the corresponding slowest relaxations. 
Figure \ref{fig:manga}
shows %, with the use of  a illustrating model,
 the relationship between
the most probable transitions from the rightmost states (blue filled curves)
and 
the evolutionary stages of local equilibria from the state.
Figure~\ref{fig:manga}(a)
describes that
the probability distribution evolves into the intra-MB local equilibrium state
in a finite time $\tau_a$, 
which is represented by the red filled curve.
Hence,
the most probable transition from the rightmost state at that time is
the transition to the minimum energy state within MB$_4$ (curved arrow).  
Figure~\ref{fig:manga}(b)
shows that
the probability distribution evolves to
the local equilibrium of 
a wider subsystem of MB$_{\{3,4\}}$, 
at  time $\tau_b$.
The most probable transition at $\tau_b$
is thus from the rightmost state to the 
minimum energy state in MB$_{\{3,4\}}$ (curved arrow).
As a result,
the probability flow from MB$_4$ to MB$_3$ is generated at  $\tau_2 \lesssim\tau_b$,
which corresponds to the relaxation mode of rate $\sim -1/\tau_2$ between
MB$_3$ and MB$_4$.
Further,
as depicted in Fig.~\ref{fig:manga}(c), 
in a global equilibration  time of $\tau_c$,
the probability distribution 
evolves to the global equilibrium (red filled curve). 
The most probable transition at $\tau_c$ is, hence,
from the rightmost state to the minimum energy state (curved arrow).
Hence,
the probability flow from MB$_{\{3,4\}}$ to MB$_{\{1,2\}}$ 
is generated at $\tau_1\lesssim \tau_c$,
which corresponds to the slowest relaxation mode of rate $\sim -1/\tau_1$
between MB$_{\{1,2\}}$ and  MB$_{\{3,4\}}$.
Here,
we have reconfirmed that there exist
switches of the most probable transitions from a state,
corresponding to 
changes in the development stages of local equilibria 
starting from the state.
At each  switching time of  $\tau=\tau_i$ ($i=1,2$),
the probability flow, which changes the local equilibrium to a  wider one,
is generated
and is  related to the slowest relaxations mode
with a relaxation rate of  $\sim -1/\tau_i$ ($i=1,2$).
Note that
in this illustrating model,
\begin{align}
\tau_a<\tau_2<\tau_b<\tau_1<\tau_c \label{eq:mangasiki}
\end{align}
holds.
Hence,
the glass-transition time $\tau_g$,
before which the inter-MB transitions are  effectively prohibited,
is given by $\tau_g=\min\{\tau_1,\tau_2,\dots\}=\tau_2$.

\subsection{{Correction formula for estimation of $\lambda_i$}\label{hosei}}
As seen in Sec.~\ref{sec:gt_tau_g},
there are small discrepancies between $\lambda_i$ and $-1/\tau_i$:
$\lambda_4\simeq -1/\tau_4$,
$\lambda_3\lesssim -1/\tau_3$,
$\lambda_2\lesssim -1/\tau_2$.
Now,
the reason is apparent.
The changes of local equilibria,
which correspond to the relaxation modes,
introduce the change of the most probable paths.
Since
the relaxation times of $-1/\lambda_i$
are followed by the structural change times $\tau_i$,
$-1/\lambda_i<\tau_i$, and  equivalently
 $\lambda_i < -1/\tau_i$, hold.
Hence, the discrepancies arise from
the lag times from the relaxations
to the local equilibrium changes.

The above discussion suggests
that
we can  obtain more accurate estimates of $\lambda_i$
from the merging process data,
by correcting the lag-time errors.
%
%For this
Toward that end,
we first evaluate the merging time $\tau$,
under the condition  that
an initial local equilibrium distribution $\vb*{p}_a$
decays to another local equilibrium distribution $\vb*{p}_b$
by a relaxation mode $\vb*{v}$.
Since $\vb*{p}_a-\vb*{p}_b \propto \vb*{v}$,
the probability distribution at $t$
is given by
$\vb*{p}(t)=\vb*{p}_b+\exp(\lambda t)(\vb*{p}_a-\vb*{p}_b)$.
Then,
the most probable path from
the maximum probability state $j_0$ of $\vb*{p}_a$
is $j_0\to j_a$ for $t<\tau$,
where $j_a$ is the second maximum probability state of $\vb*{p}_a$, 
%i.e., $(\vb*{p}_1)_{j_0}=\max\{(\vb*{p}_1)_j \mid 1\leq j \leq n\}$,
% is up to the second maximum probability state $j_1$ of $\vb*{p}_1$,
%i.e., $(\vb*{p}_1)_{j_1}=\max\{(\vb*{p}_1)_j \mid 1\leq j \leq n, j\neq j_0\}$
% for $t<\tau$,
and
it
is
$j_0\to j_b$ for $\tau<t$,
where $j_b$ is  the maximum probability state of $\vb*{p}_b$.

% up to the maximum probability state $j_2$ of $\vb*{p}_2$,
%i.e., $(\vb*{p}_2)_{j_2}=\max\{(\vb*{p}_2)_j \mid 1\leq j \leq n\}$,
% for $\tau<t$.
Hence,
at the merging time $\tau$,
$(\vb*{p}(\tau))_{j_a}=(\vb*{p}(\tau))_{j_b}$ holds.
By solving this equation for $\tau$,
we have
\begin{align}
\tau
=
-\frac1\lambda\log\left(
 1+
 \frac{(\vb*{p}_a)_{j_a}-(\vb*{p}_a)_{j_b}}{
 (\vb*{p}_b)_{j_b}-(\vb*{p}_b)_{j_a}}
\right),\label{eq:tausiki}
\end{align}
where we see that
$\tau >0$ holds,
since 
$(\vb*{p}_a)_{j_a}-(\vb*{p}_a)_{j_b}>0$,
$ (\vb*{p}_b)_{j_b}-(\vb*{p}_b)_{j_a}>0$.
From 
Eq.~\eqref{eq:tausiki},
the lag time is given by
$\tau+1/\lambda=
-1/\lambda
\left\{
\log(1+\frac{(\vb*{p}_a)_{j_a}-(\vb*{p}_a)_{j_b}}{
 (\vb*{p}_b)_{j_b}-(\vb*{p}_b)_{j_a}}
)-1\right\}
$.

By solving
Eq.~\eqref{eq:tausiki} for $\lambda$,
we obtain the formula for $\lambda$:
\begin{align}
\lambda=
-\frac1\tau\log\left(
 1+
 \frac{(\vb*{p}_a)_{j_a}-(\vb*{p}_a)_{j_b}}{
 (\vb*{p}_b)_{j_b}-(\vb*{p}_b)_{j_a}}
\right).\label{eq:lambdasiki}
\end{align}
Note here that
all values on the right-hand side of Eq.~\eqref{eq:lambdasiki}
are determined by
the metabasin-merging process data
produced by graph-based analysis.

\begin{figure}[t]
\includegraphics[width=7.5cm]{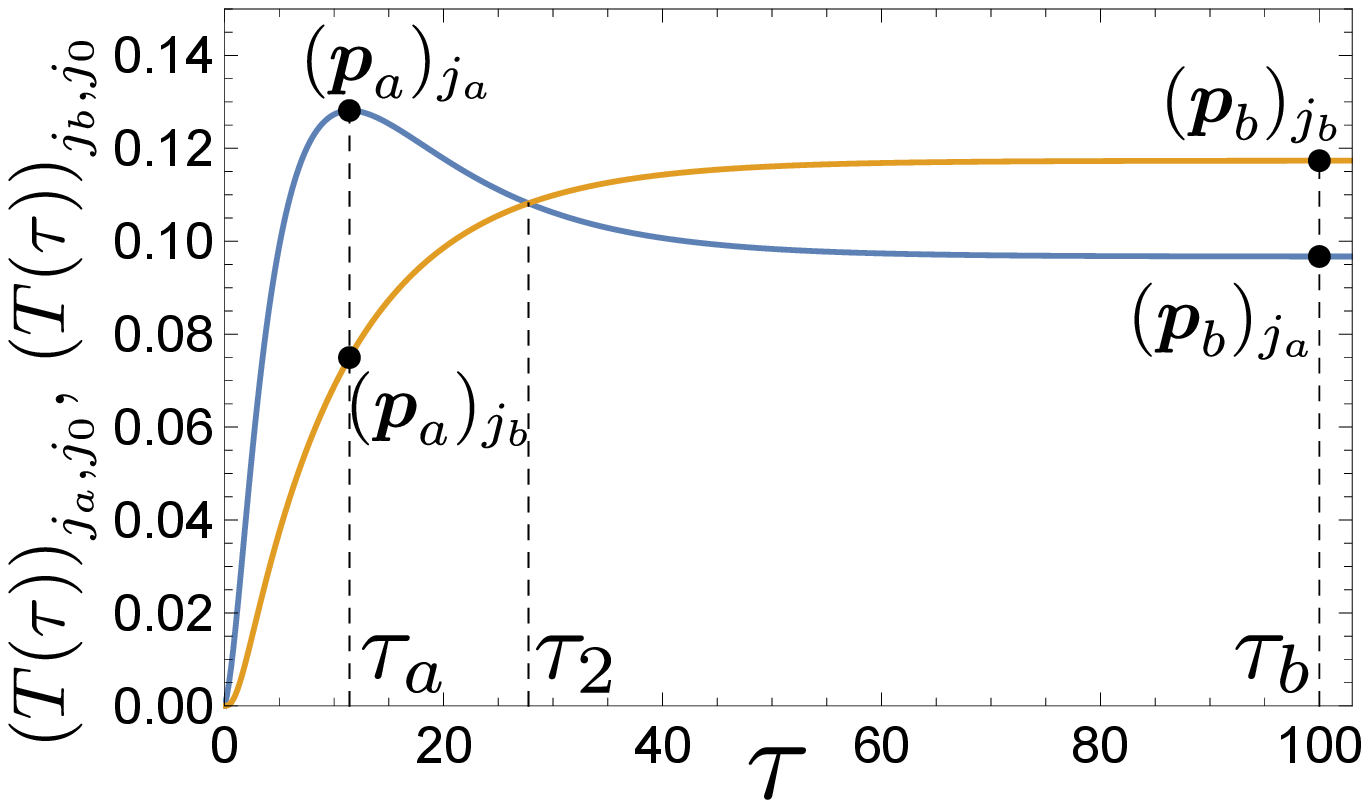}
 \caption{%
 For the evaluation of $\lambda_2$,
 $(T(\tau))_{j_a,j_0}$ and  $(T(\tau))_{j_b,j_0}$ with $j_0=23$, $j_a=39$, and
 $j_b=14$
 are plotted 
by blue and orange lines, respectively, 
 as functions of $\tau$.
The intersection at
 $\tau=\tau_2=27.76$
 corresponds to the merging process:
 MB$_{\{1,2\}}$ and MB$_{\{3,4\}}\rightarrow $MB$_{\{\{1,2\},\{3,4\}\}}$. 
The  local equilibration time $\tau_a$,
 $(\vb{p}_a)_{j_a}$,
 $(\vb{p}_a)_{j_b}$,
 $(\vb{p}_b)_{j_a}$, and
 $(\vb{p}_b)_{j_b}$,
that are  necessary for the evaluation of
Eq.~\eqref{eq:lambdasiki},
 are shown in this figure. 
 \label{fig:suitei}} 
\end{figure}

Let us examine the accuracy of Eq.~\eqref{eq:lambdasiki}
with the use of
the merging process data
of $K$
% ,
% which were discussed
in Sec.~\ref{sec:gt_tau_g}.
As shown in Sec.~\ref{sec:gt_tau_g},
the relaxation mode $\vb*{v}_2$
induces
a change of the most probable path
from
$27\to 39$ to $27 \to 14$
at $\tau_2=27.76$.
With
$j_0=27$,
$j_a=39$, and
$j_b=14$,
$(T(\tau))_{j_a,j_0}$ and 
$(T(\tau))_{j_b,j_0}$
are plotted
as functions of $\tau$
in Fig.~\ref{fig:suitei}.
They surely
have an intersection
at $\tau=\tau_2=27.76$.
$(T(\tau))_{j_a,j_0}$ takes the maximum value at
$\tau=\tau_a\equiv 11.4$ and
decreases monotonically
at $\tau>\tau_a$.
Hence,
$\tau_a$ is interpreted as the time required to reach the local equilibrium.
Hence,
in Eq.~\eqref{eq:lambdasiki}
we set
$\tau=\tau_2-\tau_a$.
Accordingly,
we set
$(\vb*{p}_a)_{j_a}=(T(\tau_a))_{j_a,j_0}=0.1281$,
$(\vb*{p}_a)_{j_b}=(T(\tau_a))_{j_b,j_0}=0.0749$,
$(\vb*{p}_b)_{j_a}=\min\{ (T(\tau))_{j_a,j_0} \mid \tau_a<\tau \leqslant \tau_b \}=0.0967$,
and
$(\vb*{p}_b)_{j_b}=\max\{ (T(\tau))_{j_b,j_0} \mid \tau_a<\tau\leqslant \tau_b \}=0.1173$ with $\tau_b=100$,
as illustrated in Fig.~\ref{fig:suitei}.
Under these conditions,
we evaluated 
Eq.~\eqref{eq:lambdasiki}
and have a result of
% resulting in
$\lambda=-0.078$.
This is
an approximate value of the exact $\lambda_2=-0.089$,
which is
much better than  the merging rate estimate of $-1/\tau_2=-0.036$.
Note here that,
Fig.~\ref{fig:suitei} 
clearly shows that
the relation $\tau_a<\tau_2<\tau_b$ holds.
Namely,
we have reconfirmed
that
Eq.~\eqref{eq:mangasiki}
holds for the four-funnel model.
\begin{table}[b]
 \caption{%
 The  $i$th relaxation rates $\lambda_i$,
 relaxation times $-1/\lambda_i$,
 merging times $\tau_i$,
 merging rates $-1/\tau_i$
 and 
 the values of Eq.~\eqref{eq:lambdasiki},   
 of the four funnel model  $K$.
 The differences between 2nd and 3rd rows
 are the lag times.}
  \label{table:fourfunelEigen}
  \centering
  \begin{tabular}{p{1.2cm}p{1.7cm}p{1.7cm}p{1.7cm}}
    \hline
    \hfil$i$\hfil  & \hfil 2\hfil  & \hfil 3\hfil  &\hfil 4\hfil\\
    \hline \hline
    \hfil$\lambda_i$\hfil  &\hfil $-0.089$\hfil  &\hfil $-0.154$\hfil &\hfil$-0.235$ \hfil\\
    \hfil$-1/\lambda_i$\hfil  & \hfil 11.29\hfil   &\hfil 6.49\hfil &\hfil 4.34\hfil \\
   \hfil $\tau_i$
   \hfil  & \hfil27.76\hfil  & \hfil 10.35\hfil &\hfil 4.20 \hfil\\
   \hfil $-1/\tau_i$
   \hfil  & \hfil$-0.036$\hfil  & \hfil $-0.0966$\hfil &\hfil $-0.238$ \hfil\\
       \hfil Eq.~\eqref{eq:lambdasiki}\hfil  & \hfil$-0.078$\hfil  & \hfil$-0.165$\hfil &\hfil $-0.227$ \hfil\\
    \hline
  \end{tabular}
\end{table}

As shown in
Table \ref{table:fourfunelEigen},
we evaluated
the values of 
$\lambda_3$ and $\lambda_4$
with the use of Eq.~\eqref{eq:lambdasiki},
from which
we see that
Eq.~\eqref{eq:lambdasiki}
generally gives
accurate approximations
of $\lambda_i$.
This means that
one merging process occurring in the most probable path graph
is effectively driven by
just one corresponding relaxation mode $\vb*{v}_i$ for $i=2,3,4$
and further that
these merging processes, as well as the relaxation modes,
are spatially separated
and 
can be
treated to be decoupled from each other.
%decouple in spaces.

\subsection{{Degenerate $\lambda_i$ cases}\label{subsubDegenerate}}
We have considered
the four-funnel model with
the random connectivity between the states,
as depicted in Fig.~\ref{fig:scg}.
Due to the randomness,
this model
has nondegenerate eigenvalues of $\lambda_i$.
However,
in particular cases, such as systems with some symmetry,
the eigenvalues $\lambda_i$ can be degenerate.
Here,
we extend our graph-based arguments
to such  degenerate $\lambda_i$ cases.

\begin{table}[b]
 \caption{
The  $i$th eigenvalues $\lambda_i$ and  the corresponding merging times $\tau_i$
 ($i=2,3,4$)
 for the  modified four funnel model $K'$ of Eq.~\eqref{mod1}.
 % are listed.
 $K'$ has the degenerate eigenvalues of  $\lambda_3=\lambda_4=-0.154$.
 % Eigenvalues other than $\lambda_4$ do not change.
 The lag times $\tau_i+1/\lambda_i>0$ are different between
 $i=3$ and $i=4$ modes,
 although they are in the same eigenspace of $\lambda=-0.154$.
 Similarly to   Table \ref{table:fourfunelEigen},
 the values of $\lambda_i$ of $K'$
 with the use of the correcting formula
Eq.~\eqref{eq:lambdasiki} were evaluated, where 
 the degeneracy of  $i=3, 4$ is recovered.
 }
  \label{modifiedEigen}
  \centering
  \begin{tabular}{p{1.2cm}p{1.7cm}p{1.7cm}p{1.7cm}}
    \hline
    \hfil$i$\hfil  & \hfil 2\hfil  & \hfil 3\hfil  &\hfil 4\hfil\\
    \hline \hline
    \hfil$\lambda_i$\hfil  &\hfil $-0.089$\hfil  &\hfil $-0.154$\hfil &\hfil$-0.154$ \hfil\\
    \hfil$-1/\lambda_i$\hfil  & \hfil 11.29\hfil   &\hfil 6.49\hfil &\hfil 6.49\hfil \\
   \hfil$\tau_i$\hfil  & \hfil27.76\hfil  & \hfil 10.35\hfil &\hfil 7.4 \hfil\\
   \hfil$-1/\tau_i$\hfil  & \hfil$-0.0360$\hfil  & \hfil $-0.0966$\hfil &\hfil $-0.135$ \hfil\\
    \hfil Eq.~\eqref{eq:lambdasiki}\hfil  & \hfil$-0.0473$\hfil  & \hfil $-0.165$\hfil &\hfil $-0.165$ \hfil\\   
   \hline
  \end{tabular}
\end{table}

First,
we introduce a degenerate model by modifying the four-funnel model.
To this end,
recall that
the spectral representation of
the transition rate matrix $K$ of the four-funnel model
is given by
\begin{align}
 K&=V\Lambda V^{-1}\\
 V&=\left[\vb*{v}_1,\vb*{v}_2,\vb*{v}_3,\vb*{v}_4,\dots  \right],\\
 \Lambda&=\text{diag}(\lambda_1,\lambda_2,\lambda_3,\lambda_4,\dots).
\end{align}
Now, we consider
the degenerate matrix $K'$
that is obtained by changing the value of $\lambda_4$ to the value
of $\lambda_3$ in $K$,
whose
spectral representation
is
given by
\begin{align}
 K'&=V\Lambda'V^{-1}\label{mod1}\\
  \Lambda'&=\text{diag}(\lambda_1,\lambda_2,\lambda_3,\lambda_3,\dots). 
\end{align}
The most probable path graphs of
 the transition probability matrix $T'(\tau)=\exp (\tau K')$
 were examined,
 where
 we found
 the same merging processes as depicted in Fig.~\ref{fig:tau4.2}.
 However,
the merging times $\tau_i$  were changed
as shown in Table~\ref{modifiedEigen}.
From Table~\ref{modifiedEigen},
we see that
the doubly degenerate eigenvalues of $\lambda_3=\lambda_4$
correspond to
the resolved merging times $\tau_3>\tau_4$,
which
means
that
lag times, $\tau_i+1/\lambda_i$,
from the relaxation times $-1/\lambda_i$
to the merging times $\tau_i$
are different between $i=3$ and $i=4$ modes.
As shown in  Table \ref{modifiedEigen},
 the values of $\lambda_i$ of $K'$
 are estimated
 with the use of the correcting formula of 
Eq.~\eqref{eq:lambdasiki}.
 Table \ref{modifiedEigen}
 clearly shows that
 the degeneracy of  $\lambda_3=\lambda_4$ is revealed
 by this estimation.
%  in the estimated $\lambda_i$ values.

Assuming here that
the system had
the same lag times of $i=3$ and $i=4$,
then
$\tau_3=\tau_4$  would hold
and 
the two merging processes
would simultaneously occur at $\tau=\tau_3$:
one was 
MB$_1(1\Leftrightarrow 2)$ and  MB$_2(14\Leftrightarrow 15)\longrightarrow$
MB$_{\{1,2\}}(2\Leftrightarrow 1\leftarrow 14\leftarrow 15)$,
which would induce the probability flow
corresponding to the eigenvectors $\vb*{v}_4$,
and
the other
was
MB$_3(27\Leftrightarrow 28)$ and 
MB$_4(39\Leftrightarrow 40)\longrightarrow$
MB$_{\{3,4\}}(28\Leftrightarrow 27\leftarrow 39\leftarrow 40)$,
which would induce the probability flow
corresponding to  $\vb*{v}_3$.
Hence,
we were able to extract the relaxation modes $\vb*{v}_3$, $\vb*{v}_4$
by merging graph analysis
without any change,
in this degenerate lag-time case.

\begin{table}[b]
 \caption{
The  $i$th eigenvalues $\lambda_i$ and  the corresponding merging times $\tau_i$
 ($i=2,3,4$)
 for the  modified four funnel model $K''$ of Eq.~\eqref{mod2}.
 % are listed.
 $K''$ has the degenerate eigenvalues of
 $\lambda_2=\lambda_3=\lambda_4=-0.089$.
 % Hence, $-1/\lambda_i=11.29 $ and
 The lag times $\tau_i+1/\lambda_i>0$ are different among these modes,
 although they are in the same eigenspace of $\lambda=-0.089$.
 Similarly to  Tables \ref{table:fourfunelEigen} and \ref{modifiedEigen},
 we estimate the values of $\lambda_i$ of $K''$
 with %the use of the correcting formula
Eq.~\eqref{eq:lambdasiki}, where 
 the degeneracy of  $i=2, 3, 4$ is almost recovered.
 }
  \label{modifiedEigen2}
  \centering
  \begin{tabular}{p{1.2cm}p{1.7cm}p{1.7cm}p{1.7cm}}
    \hline
    \hfil$i$\hfil  & \hfil 2\hfil  & \hfil 3\hfil  &\hfil 4\hfil\\
    \hline \hline
    \hfil$\lambda_i$\hfil  &\hfil $-0.089$\hfil  &\hfil $-0.089$\hfil &\hfil$-0.089$ \hfil\\
    \hfil$-1/\lambda_i$\hfil  & \hfil 11.29\hfil   &\hfil 11.29\hfil &\hfil 11.29\hfil \\
    \hfil$\tau_i$\hfil  & \hfil20.2\hfil  & \hfil 15.4\hfil &\hfil 13.0 \hfil\\
    \hfil$-1/\tau_i$\hfil  & \hfil$-0.0769$\hfil  & \hfil$-0.0649$\hfil &\hfil $-0.0495$ \hfil\\
    \hfil Eq.~\eqref{eq:lambdasiki}\hfil  & \hfil$-0.099$\hfil  & \hfil$-0.103$\hfil &\hfil$-0.111$\hfil\\
    \hline
  \end{tabular}
\end{table}

Finally,
we consider a triply degenerate matrix $K''$,
that is obtained by changing the values of $\lambda_4$ and $\lambda_3$ to the value of $\lambda_2=-0.089$ in $K$, 
whose transition rate matrix is given by
\begin{align}
 K''&=V\Lambda''V^{-1}\label{mod2}\\
 \Lambda''&=\text{diag}(\lambda_1,\lambda_2,\lambda_2,\lambda_2,\dots).
 \end{align}
 
 The merging times are listed in Table~\ref{modifiedEigen2},
 which clearly shows that
the lag times, $\tau_i+1/\lambda_i$,
are different among  $i=2$, $3$ and $4$ modes
in this triply degenerate case too.
 
\begin{figure}[t]
 \includegraphics[width=6cm]{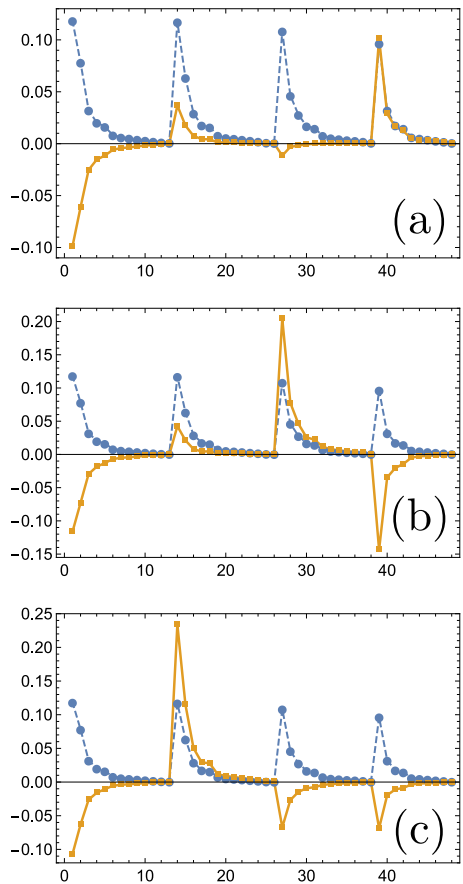}  
\caption{%
%In Figs. (a), (b), and (c), 
Eigenvectors   
  of transition rate matrix $K''$ for the modified four-funnel model
  of Eq.~\eqref{mod2}:
 (a) $\vb*{v}'_4$,  (b) $\vb*{v}'_3$, and  (c) $\vb*{v}'_2$. 
In each plot,
the equilibrium, $\vb*{v}_1$,
is also plotted using circles connected with a dashed line for comparison.
Here, 
$\vb*{v}'_k$ ($k\geqslant 2$) are scaled 
such that
the components satisfying $(\vb*{v}'_k)_i<0$ approximately agree with $-(\vb*{v}_1)_i$.
\label{fig7}
} 
 \end{figure}

 At $\tau_4=13.0$,
 the most probable path from 39 changes
 from $39\to 40$ to $39\to 1$,
 so that
 MB$_1$ and MB$_4$ merged into MB$_{\{1,4\}}$, 
 which indicates the probability flow from  MB$_4$ to MB$_1$.
The corresponding relaxation mode 
$\vb*{v}'_4=\vb*{v}_2
+\vb*{v}_3-\vb*{v}_4$
does exist
in the eigenspace of $\lambda=-0.0089$,
as shown in Fig.~\ref{fig7}(a).
At $\tau_3=15.4$,
the most probable path from 27 changes
from  $27\to 28$ to $27\to 1$,
so that
MB$_3$ and  MB$_{\{1,4\}}$ next merged into 
MB$_{\{\{1,4\},3\}}$. 
Hence, the probability flow
from MB$_3$ to MB$_{\{1,4\}}$ is expected.
The corresponding relaxation mode
$\vb*{v}'_3=1/2 \vb*{v}_2+1/2 \vb*{v}_3+\vb*{v}_4$
exists in the eigenspace,
as shown in Fig.~\ref{fig7}(b),
from which
we see that
the excess probability stored in MB$_3$
is transported into
MB$_1$ and MB$_4$ by $\vb*{v}'_3$.
Lastly,
at 
$\tau_2=20.2$,
the most probable path from 14
changes from $14\to 15$ to $14\to 1$,
so that
MB$_2$ and 
MB$_{\{\{1,4\},3\}}$ merge into 
MB$_{\{\{\{1,4\},3\},2\}}$.
Hence,
the probability flow from
MB$_{2}$
to
MB$_1$,
MB$_3$, and
MB$_4$
is expected.
The corresponding  relaxation mode
is 
$\vb*{v}'_2=\vb*{v}_2-2\vb*{v}_3$,
as shown in Fig.~\ref{fig7}(c).
In short,
we have confirmed
that
in the case of triply degenerate eigenvalues,
the three merging processes at $\tau=\tau_4,\tau_3,\tau_2$
exist due to the different lag times.
These merging processes 
correspond, respectively, to
three linearly independent  eigenvectors of 
$\vb*{v}'_4$,
$\vb*{v}'_3$, and
$\vb*{v}'_2$,
in the same eigenspace of $\lambda$.

 As shown in  Table \ref{modifiedEigen2},
 the values of $\lambda_i$ 
 are estimated for  $K''$,
 with the use of the correcting formula of 
Eq.~\eqref{eq:lambdasiki}.
 Table \ref{modifiedEigen2}
 clearly shows that
 the degeneracy of  $i=2, 3, 4$ is almost reconstructed
 in the estimated $\lambda_i$ values.
Hence, in this triply degenerate case,
each relaxation mode corresponds to each merging process,
which enables us to 
determine 
the values of $\lambda$ accurately,
 with the use of Eq.~\eqref{eq:lambdasiki}.

Here,
assuming again that
the three lag times were the same,
then,
all MBs merged into
MB$_{\{\{\{1,4\},3\},2\}}$
at a certain $\tau$,
where
the most probable paths
simultaneously change
from $39\to 40$ to $39\to 1$,
from $27\to 28$ to $27\to 1$,
and
from $14\to 15$ to $14\to 1$.
In this case, too,
we could extract
the three linearly independent eigenvectors
of
$\vb*{v}'_4$,
$\vb*{v}'_3$, and
$\vb*{v}'_2$,
in the eigenspace of $\lambda$,
by
resolving the accumulating merging
into 
the above three separated mergings.
Of course,
we might resolve the simultaneous merging
into 
other separated mergings:
e.g.,
MB$_1$ and 
MB$_3$ merged into MB$_{\{1,3\}}$;
MB$_{\{1,3\}}$ and 
MB$_2$ merged into 
MB$_{\{\{1,3\},2\}}$;
and
MB$_{\{\{1,3\},2\}}$ and
MB$_4$ merged into
MB$_{\{\{\{1,3\},2\},4\}}$ after that.
In this separation,
we would obtain another set of
linearly independent
eigenvectors of
$\vb*{v}''_2$,
$\vb*{v}''_3$, and
$\vb*{v}''_4$,
in the same eigenspace of $\lambda$.

In summary,
we have found the correspondences between
the merging processes and the eigenvectors
in the case of the  eigenspace of $\lambda$ with multiple degeneracies.
A single merging process corresponds to
an eigenvector in the eigenspace of $\lambda$.
Generally,
the lag times from the relaxation times $-1/\lambda_i$
to the merging times $\tau_i$
vary from eigenvector to eigenvector.
Thus,
the quasi-degenerate merging rates, $-1/\tau_i$, are resolved.
% and
% quasi-degenerate.
If the lag times are equal,
separated reroutings of the most probable paths
are postulated in the merging process,
from which we can extract
the corresponding eigenvectors from the eigenspace of $\lambda$.
These eigenvectors
carry the probability flows induced by the postulated merging processes.
Hence,
we can extract
the eigenvalues and the eigenvectors
that correspond to the elemental merging processes,
from any degenerate systems,
within the error of lag times.
Furthermore,
with the use of Eq.~\eqref{eq:lambdasiki},
we can extract the accurate values of $\lambda_i$,
which are free from the lag-time errors,
from the merging process data of degenerate, as well as nondegenerate,
$\lambda_i$ systems.

\section{conclusion}
In this paper,
we have considered
the  structural changes of 
the most probable path graphs
of $T(\tau)$.
The parameter $\tau$ is  a coarse-graining parameter 
in that
the  modes relaxing faster than $1/\tau$ in rate
are neglected from $T(\tau)$.
As $\tau$
is increased,
the most probable path graphs are frequently reconnected, 
where there exists a specific glass-transition time $\tau_g$,
which divides $\tau$ into two qualitatively different regions.

For $\tau<\tau_g$,
the members of the metabasins
(i.e., the connected graph components)
remain almost unchanged,
and 
only the intra-metabasin local equilibria can be attained.
We have confirmed
that 
not only for transition rate matrices $K$
but also for transition probability matrices $T(\tau)$
the metabasins 
are suitable bases
both for coarse-graining and for renormalization procedures in Ref.~\cite{rg},
 since these procedures  are not sensitive to
 the values of  $\tau$ when $\tau<\tau_g$.

On the other hand,
for $\tau \geqslant \tau_g$,
the inter-metabasin reconnections of attracting cycles,
which lead to the mergings of metabasins, 
occur three times.
For each value of $\tau$ at which metabasins merge with each other,
there exists an eigenvalue around the rate of $-1/\tau$,
and 
the corresponding eigenvector clearly shows
that
the relaxation process corresponds exactly
to the merging process of metabasins.

{
In conclusion,
we have revealed
that
the relaxation properties can be extracted
via
analyzing structural changes of
the most probable path graphs of $T(\tau)$.
The advantages of our graph-based method are as follows:
(a)
In our method,
metabasins are extracted visually directly from the most probable path graphs.
In contrast,
%to this,
%Compared to this,
in the other widely-used methods, such as the Perron cluster algorithm \cite{BPE}, 
some processing of diagonalizations and linear superpositions
 is necessary for  extracting  metabasins.
% transition  matrices are  diagonalized,
% and then
% linear superposition processing is applied to the obtained eigenvectors.
%
(b)
From the merging of metabasins of the most probable path graphs at $\tau_i$,
we can evaluate  the slowest relaxation rates $\lambda_i$ as about $-1/\tau_i$
 and the corresponding eigenvectors as the probability flows
 between just merging metabasins.
 Furthermore,
 with the use of Eq.~\eqref{eq:lambdasiki},
which
 corrects the lag times between  $-1/\lambda_i$ and $\tau_i$,
 one can evaluate the  value of $\lambda_i$ with high accuracy
 from the merging process.
 (c) These method developed in this paper are available for
 a  wide range of kinetic systems with degenerate, as well as
 non-degenerate, relaxation rates.
 % ranging from non-degenerate to degenerate rate kinetic system.
% the degenerate , as well as non-degenerate, relaxation rate systems.
}

We remark finally that
one can start the metabasin analysis developed in this paper
only with the information about the most probable transitions $i\to j$.
The states $i$ and  the transition probability matrices $T(\tau)$,
required for this analysis, 
can be estimated by various clustering methods
both from simulation datasets and
from  experimental datasets \cite{BPE,hummer3}.
Therefore, even if a kinetic system was very complicated,
it would be relatively easily to extract the information about 
the most probable transitions and thus  the slowest relaxation modes
from the transition data via the metabasin analysis.
Hence,
we hope that
this simple graph-based analysis, developed in this work,
will be applied to a wide range of
realistic kinetic systems
for extracting 
the slowest relaxation modes
via 
experimentally or numerically accessible %obtainable
transition probability matrices.

\begin{acknowledgments}
Y.~S.~and T.~O.~are supported by 
Grant-in-Aid for Challenging Exploratory Research
(Grant No. JP15K13539) 
from the Japan Society for the Promotion of Science. 
 T.~O.~expresses gratitude to Naoto Sakae
 and
 Kiyofumi Okushima 
for 
enlightening discussions and continuous encouragement.
The authors are very grateful to Shoji Tsuji and Kankikai 
for the use of their facilities at Kawaraya during this study.

\end{acknowledgments}

\end{document}